\documentclass{aa}
\usepackage{bm}
\usepackage{txfonts}
\usepackage[T1]{fontenc}
\usepackage[colorlinks=true,citecolor=blue,linkcolor=blue,urlcolor=red,breaklinks]{hyperref}                         
\usepackage{amssymb}                    
\usepackage{amsmath}                    
\usepackage{enumerate}                  
\usepackage{float}                      
\usepackage{journals}                   
\usepackage{natbib}                     
\usepackage{textcomp}                   
\usepackage{footnote}                   
\usepackage{textcomp}                   
\usepackage{tabularx}                   
\usepackage{multirow}                   
\usepackage{booktabs}                   
\usepackage[font=small,labelfont=bf]{caption} 
\usepackage{framed}                             
\usepackage{bm}                                         
\usepackage{graphicx}                   
\usepackage{enumitem}                   
\usepackage{subfigure}                  

\usepackage{threeparttable} 

\def\bfref{} 
\def\bmref{} 

\def\bf2ref{}   
\def\bm2ref{}   

\def\bffinal{}   
\def\bmfinal{}   

\begin{document}


\title{The LOFAR search for radio pulsars and fast transients in M33, M81 \& M82}

\author{K.~Mikhailov\inst{\ref{inst1},\bffinal{\ref{inst2}}}\and J.~van~Leeuwen\inst{\ref{inst2},\ref{inst1}}}

\institute{Anton Pannekoek Institute for Astronomy, University of Amsterdam, Science Park 904, 1098 XH Amsterdam, The Netherlands\\\email{K.Mikhailov@uva.nl}\label{inst1} 
\and
ASTRON, the Netherlands Institute for Radio Astronomy, Postbus 2, 7990 AA, Dwingeloo, The Netherlands\\\email{leeuwen@astron.nl}\label{inst2}
}
\date{Received \dots; accepted \dots}

 
\abstract
{The radio pulsar and rotating radio transient populations are only known in \bfref{and near} the Milky Way. Investigating such populations in other galaxies requires deep pulsar and transient searches. We performed 4-h radio observations of nearby galaxies M33, M81 and M82 with LOFAR.}
{Our main purpose was to characterise the bright end of the pulsar population in other galaxies, and compare it to that of the Milky Way.}
{We searched for extragalactic radio pulsars through a periodic-pulse search, and for sporadic fast radio transients through a single-pulse search. We coherently combined \bffinal{at most 23} LOFAR Core High-Band Antenna (HBA) stations and covered M33, M81, and M82 in their entirety using multiple tied-array beams.}
{No pulsating sources or single pulses were found. We have, therefore established stricter limits on the extragalactic pulsar flux density at lower frequencies than those obtained in previous Arecibo\bffinal{, GBT,} and WSRT searches.}
{We conclude that in nearby galaxies M33, M81, and M82 there are no pulsars \bfref{shining toward Earth} with pseudo luminosities greater than \bffinal{a few} times that of the brightest pulsars in our Milky Way.}


\maketitle
%

\section{Introduction}\label{sec1}

Radio pulsars~\citep{Hewish-1968} are rapidly rotating neutron stars whose physical facets differ strongly from those encountered on Earth: pulsars have, for example, magnetic fields that are twelve orders of magnitude stronger than those of Earth; yet they are three orders of magnitude smaller. There are now more than 2500 pulsars known within our Galaxy, including its globular clusters\footnote{\url{http://www.atnf.csiro.au/people/pulsar/psrcat} (catalogue version 1.54)}~\citep{Manchester-2005}. 

Normal radio pulsars emit radio waves along the field lines over their magnetic poles; as the pulsar continuously spins and radiates, its radio emission is ultimately caught by radio telescopes as periodic pulses~\citep[see][for a general overview of pulsar radio emission]{LorimerKramer-2004}. Such pulses can be \bfref{integrated} to get a \bfref{higher signal-to-noise (S/N)} pulse profile. Rotating radio transients~\citep[RRATs,][]{McLaughlin-2006}, in contrast, emit only occasionally; \bfref{these sources are more easily found in single pulse searches than in periodicity searches}~\citep{Cordes-2003, Keane-2011}. These thus require a more detailed analysis of individual pulses. Similarly, it might be possible to receive extremely bright giant pulses from relatively young pulsars in distant galaxies, another benefit of single-pulse searches~\citep{McLaughlin-2003}. Catching single pulses is, however, challenging as the astronomical data is usually contaminated with radio frequency interference (RFI). Such detrimental signals can be produced by terrestrial sources in a frequency range coincident with that of the observation. 

\begin{table*}[t]
\centering
\caption{Past surveys dedicated to \bfref{the periodicity (PS) and single-pulse (SPS)} pulsar search\bfref{es} in nearby galaxies. \bfref{Tabulated are: galaxy, its distance, the telescope, central frequency, bandwidth in MHz, sampling time in $\bmref{\mu}$s, total dwell time in hours, and the obtained average (for PS), and peak (for SPS) flux densities in Jy.}}
\begin{threeparttable}
\scalebox{0.595}{
\begin{tabular}{c c c c c c c c c c}
\hline\hline
Galaxy & $d$ & Telescope & $F_\textrm{cntr}$ & $\Delta\nu$ & $T_\textrm{samp}$ & $T_\textrm{int}$ & $S_\textrm{min}$\,[PS] & $S_\textrm{peak}$\,[SPS] & Reference \\
 & (Mpc) & & (MHz) & (MHz) & ($\bmref{\mu}$s) & (hrs) & (Jy) & (Jy) & \\
\hline
\multirow{2}{*}{M33} & \multirow{2}{*}{0.84} & \multirow{2}{*}{Arecibo} & 430 & 10 & 102.4 & 3.0 & $\bmref{0.2\times10^{-3}}$ & \bfref{0.5} & 1 \\
                     &  &  & 1440 & 100 & 64 (100) & 2.0 (3.0) & $\bmref{5.3\times10^{-6}}$ & \bfref{0.1} & 2 \\[0.3cm]
NGC253, NGC300, Fornax & 3.0, 2.0, 16.9 & \multirow{2}{*}{Parkes} & \multirow{2}{*}{435} & \multirow{2}{*}{32} & \multirow{2}{*}{420} & \multirow{2}{*}{3.0} & \multirow{2}{*}{\bf2ref{(\dots)}} & \multirow{2}{*}{$\bmref{0.09}$} & \multirow{2}{*}{1} \\
NGC6300, NGC7793 & 16.9, 3.5 & & & & & & & & \\[0.3cm]
Leo I (dSph galaxy) & 0.25 & \multirow{2}{*}{GBT} & 350 & 100 & 81.92 & 20.0 & $\bmref{\sim2.0\times10^{-4}}$ & $\bmref{\sim0.04}$ & 3 \\
IC 10 & 0.66 & & 820 & 200 & 204.8 & 16.0 & $\bmref{1.5\times10^{-5}}$ & \bfref{0.02} & 4 \\[0.3cm]
Willman 1, Boo dw, UMi dw & 0.046, 0.062, 0.075 & \multirow{6}{*}{GBT (Arecibo)} & \multirow{6}{*}{820 (327)} & \multirow{6}{*}{50 (50)} & \multirow{6}{*}{81.92 (128)} & 3.3, 4.4, 5.3 & \multirow{6}{*}{\bf2ref{(\dots)}} & \bfref{0.5 (1.54), 0.3 (1), 1.6 (5), each }$\bmref{\times10^3}$ & \multirow{6}{*}{5} \\
Dra dw, Scl dw, Sex dw & 0.078, 0.093, 0.099 & & & & & 3.6, 2.8, 3.25 & & \bfref{2.3 (7), 2.3 (7.1), 2.3 (7.1), each }$\bmref{\times10^3}$ & \\
CVn I, Leo I-II, UMa II & 0.23, 0.28, 0.23, 0.25 & & & & & 6.2, 21.9, 10.9, 2.0 & & \bfref{4 (12), 6 (19), 4 (12), 19 (57), each }$\bmref{\times10^3}$ & \\
And II, III, VI, XI-XIV & 0.9 & & & & & 1.4, 1.1, 2.5, 3.9, 3.7, 3.0, 2.0 & & \bfref{64 (198), each }$\bmref{\times10^3}$ & \\
Leo T, Leo A & 0.4, 0.78 & & & & & 4.2, 6.1 & & \bfref{13 (40), 44 (136), each }$\bmref{\times10^3}$ & \\
IC 1613, LGS 3, Peg dw & 0.9, 0.93, 0.93 & & & & & 3.5, 3.9, 2.5 & & \bfref{63 (195), 67 (207), 67 (207), each }$\bmref{\times10^3}$ & \\[0.3cm]
M31 & 0.8 & WSRT & 328 & 10 & 409.6 & 32.0 & $\bmref{0.3\times10^{-3}}$ & \bfref{2.8} & 6 \\
\hline
\end{tabular}
}
\tablebib{
(1)~\citet{McLaughlin-2003}; (2)~\citet{Bhat-2011}; (3)~\citet{RubioHerrera-2013}; (4)~\citet{Noori-2014}; \\ 
(5)~\citet{Vlad-2013} and Kondratiev~\citetext{priv. comm.}; (6)~\citet{Rubio-Herrera-2013}.
}
\end{threeparttable}
\label{Surveys}
\end{table*}

Several other factors usually play important roles in the apparent brightness of the pulsar signal. First off, the distance $d$ to the source clearly affects the pulsar flux density: $S\,\bmref{\propto}\,d^{-2}$. Furthermore, after the pulsar wave front has travelled through the \bfref{inter}stellar medium,  the lower frequencies arrive later than higher frequencies (dispersion, $t_\mathrm{arrival}\,\bmref{\propto}\,\nu^{-2}$). Multi-path propagation also introduces a delayed power-law tail in the integrated pulse profile (scattering, $\tau_\mathrm{scatter}\,\bmref{\propto}\,\nu^{-4}$). 

The discovery of new pulsars, either periodic or transient, provides better insights into, for example the source birth rate, the progenitor populations, the spatial and flux density distributions. The subsequent timing of the most stable radio pulsars can lead to very good tests of space-time curvature that probe gravitational effects and of supranuclear density which can establish well-fitted equations of neutron star state. Moreover, the study of frequency-dependent scattering and dispersion \bfref{can result in a more detailed understanding of the content and density profile of the free electrons in the interstellar medium~\citep{Cordes-2002}}.

For the extragalactic search, apart from the analysis of intergalactic matter, the presence of pulsars in other galaxies allows us to establish the link between the galaxy evolution and the pulsar population synthesis there. From such a link we can infer whether different galactic progenitors create different neutron star populations.

There have been numerous pulsar discoveries among relatively close stellar overdensities, for example globular clusters\footnote{\url{http://www.naic.edu/~pfreire/GCpsr.html}}~\citep[see][for a review]{Camilo-2005}, \bfref{and nearest neighbour galaxies of the Local Galactic Group:} the Small and Large Magellanic Clouds~\citep{Crawford-2001, Ridley-2013}; but also non-detections towards, for example, dwarf spheroidal galaxies~\citep{RubioHerrera-2013}. In addition, several efforts have been made to capture periodic as well as single pulse signals from nearby galaxies, with different telescopes. Unfortunately none of these searches discovered new sources. For the deepest of such  past surveys, we list the frequencies and sensitivities in Table~\ref{Surveys}. These frequencies range from 328-1440 MHz. \bfref{Based on the facts that at lower frequencies pulsar beams get wider and pulsar fluxes get higher~\citep{Stappers-2011}, }there is a possibility that deep, lower-frequency surveys could find pulsars that these past efforts have missed.

\begin{table*}[ht!]
\centering
\caption{LOFAR Observations of nearby galaxies M33, M81, and M82. For tied-array rings (TARs) only the coordinates of the central tied-array beam (TAB) are provided.}
\begin{tabular}{c | c | c | c | c | c | c}
\hline\hline
ObsID & Galaxy & Angular size & Dwell time & \textnumero~TARs\,/\,TABs & RA & DEC \\
 & & (arcmin) & (hrs) & & (J2000) & (J2000) \\
\hline
L274117 & M33 & 70.8 $\times$ 41.7 & \multirow{5}{*}{4} & 5 TARs & 01:33:50.90 & +30:39:35.8 \\ \cline{1-3} \cline{5-7}
\multirow{4}{*}{L342964} & M81 & 26.9 $\times$ 14.1 & & 3 TARs & 09:55:33.19 & +69:03:55.0 \\ \cline{2-3} \cline{5-7}
                & \multirow{3}{*}{M82} & \multirow{3}{*}{11.2 $\times$ 4.3} & & 1 TAB & 09:56:06.10 & +69:41:15.0 \\ \cline{5-7} 
                &                                               & & & 1 TAB & 09:55:52.20 & +69:40:47.0 \\ \cline{5-7} 
                &                                               & & & 1 TAB & 09:55:38.75 & +69:40:20.0 \\ \hline
\end{tabular}
\label{Obs}
\end{table*}

The LOw Frequency Array~\citep[LOFAR,][]{Haarlem-2013} is a radio telescope capable of tracking several nearby galaxies, at long wavelengths. The LOFAR high-band antenna's (HBA, 120 -- 240 MHz) frequency range approximately coincides with the peak of pulsar flux density distribution~\citep[100 -- 200\,MHz,][]{Stappers-2011}, which is an advantage for a pulsar search. Compared to previous low-frequency pulsar searches, further improvements such as the much increased bandwidth ($\approx 70\,\mbox{MHz}$), the high core gain ($\approx 8.8\,\mbox{K/Jy}$), and the relatively low antenna temperature~\citep[$\approx 160\,\mbox{K}$, see Eq. 3 from][]{Leeuwen-2010} mean that LOFAR can, on the one hand, find many new interesting pulsars~\citep[see][]{Coenen-2014}, and on the other hand, better investigate their fundamental emission peculiarities~\citep[see][]{Stappers-2011}. In general, pulsar beams are thought to be wider at lower frequencies, giving LOFAR \bfref{the advantage in capturing radio pulses} over higher-frequency surveys. Finally, the LOFAR multi-beaming allows for efficient, deep integrations \bfref{at a sensitivity and field of view unparalleled in the world}. The main downside of searching for pulsars with LOFAR is that dispersion and scattering degrade the inherently sharp pulsar peaks.
 
In this paper, we describe the searches we have performed with LOFAR for pulsars in spiral Triangulum Galaxy M33 (distance $d\approx 0.73-0.94\,\mbox{Mpc}$), Bode's Galaxy M81 ($d\approx 3.50-3.74\,\mbox{Mpc}$), and starburst Cigar Galaxy M82 ($d\approx 3.5-3.8\,\mbox{Mpc}$).  Given their northern positions (see Table~\ref{Obs}), these three galaxies can be well targeted with LOFAR, as the projected effective area of LOFAR's fixed dipoles, and thus the sensitivity, is the highest close to zenith~\citep[see][]{Leeuwen-2010}.

The origin and current environment of a galaxy inevitably affect its star-formation rate and mass-energy distribution. As a result, the pulsar population in nearby galaxies can be different from that of our Milky Way, and a comparison between this population and that of our Milky Way is interesting.

\bfref{Furthermore, measuring a dispersion measure (DM) toward an extragalactic pulsar immediately indicates the total amount of electron matter along the line of sight. Once an ensemble of such pulsars are found in a certain galaxy, one may be able to disentangle the host and interstellar components, and measure the free electron density of the intergalactic matter, a quantity highly interesting for fast radio burst (FRB) studies~\citep{Spitler-2016}\bffinal{, for example.}}

Our paper is organised as follows: Section~\ref{sec2} outlines the general characteristics of observational pointings; Section~\ref{sec3} describes the actual search procedure. The search results and sensitivity estimations are given in Section~\ref{sec4}, and we present our conclusions in Section~\ref{sec5}.

\section{Observations}\label{sec2}

The ability of LOFAR to produce multiple highly sensitive tied-array beams (TABs) provides an opportunity to fully cover a targeted nearby galaxy and therefore carry out a deep search over its entire area. Our first targeted observation was directed at the M33 galaxy, while the second captured both M81 and M82. 

\begin{figure}[t!]
\centering\subfigure[M33 galaxy]{
\includegraphics[width=0.7\linewidth]{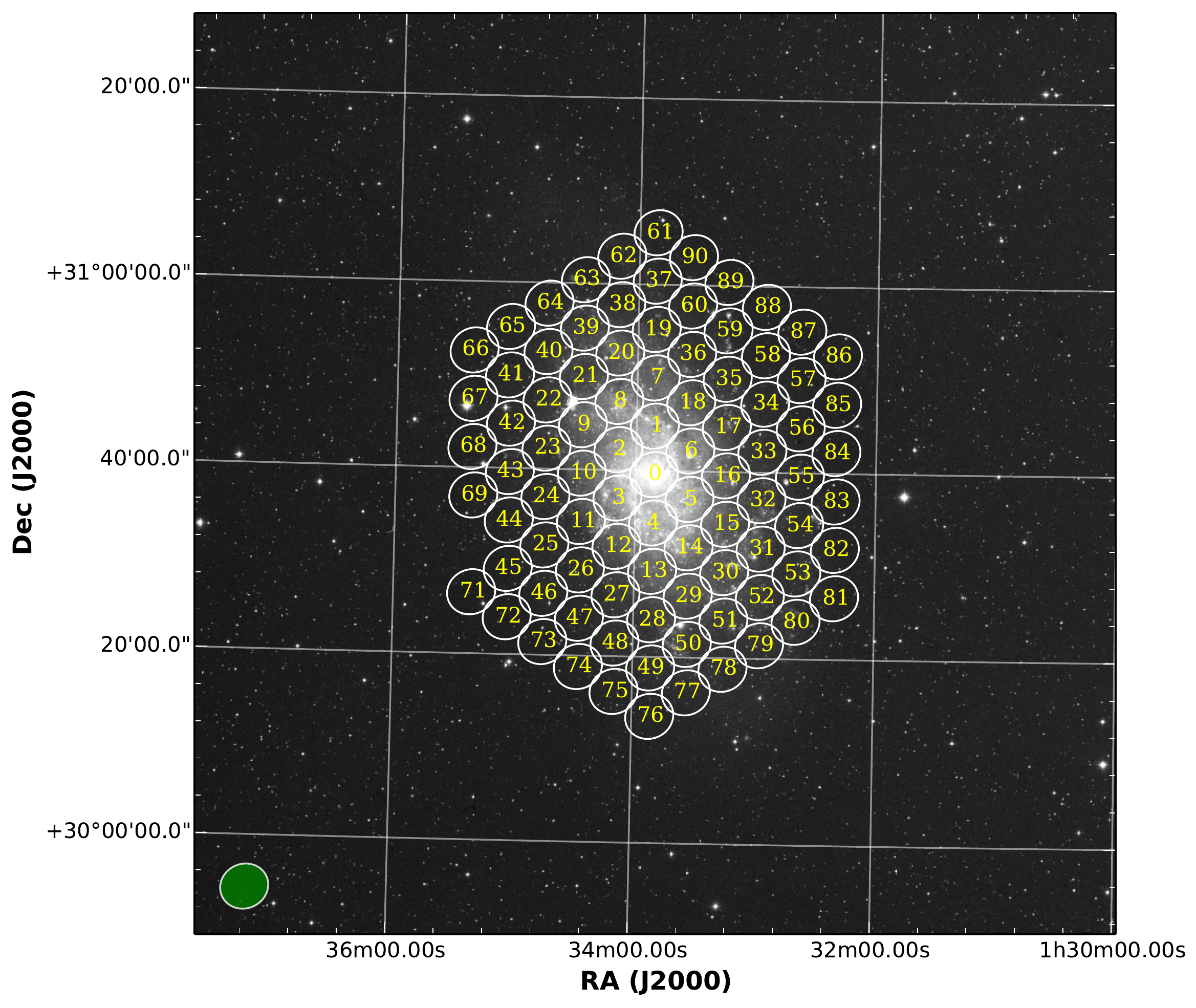}\label{m33}
}
\vfill
\centering\subfigure[M81 and M82 galaxies]{
\includegraphics[width=0.7\linewidth]{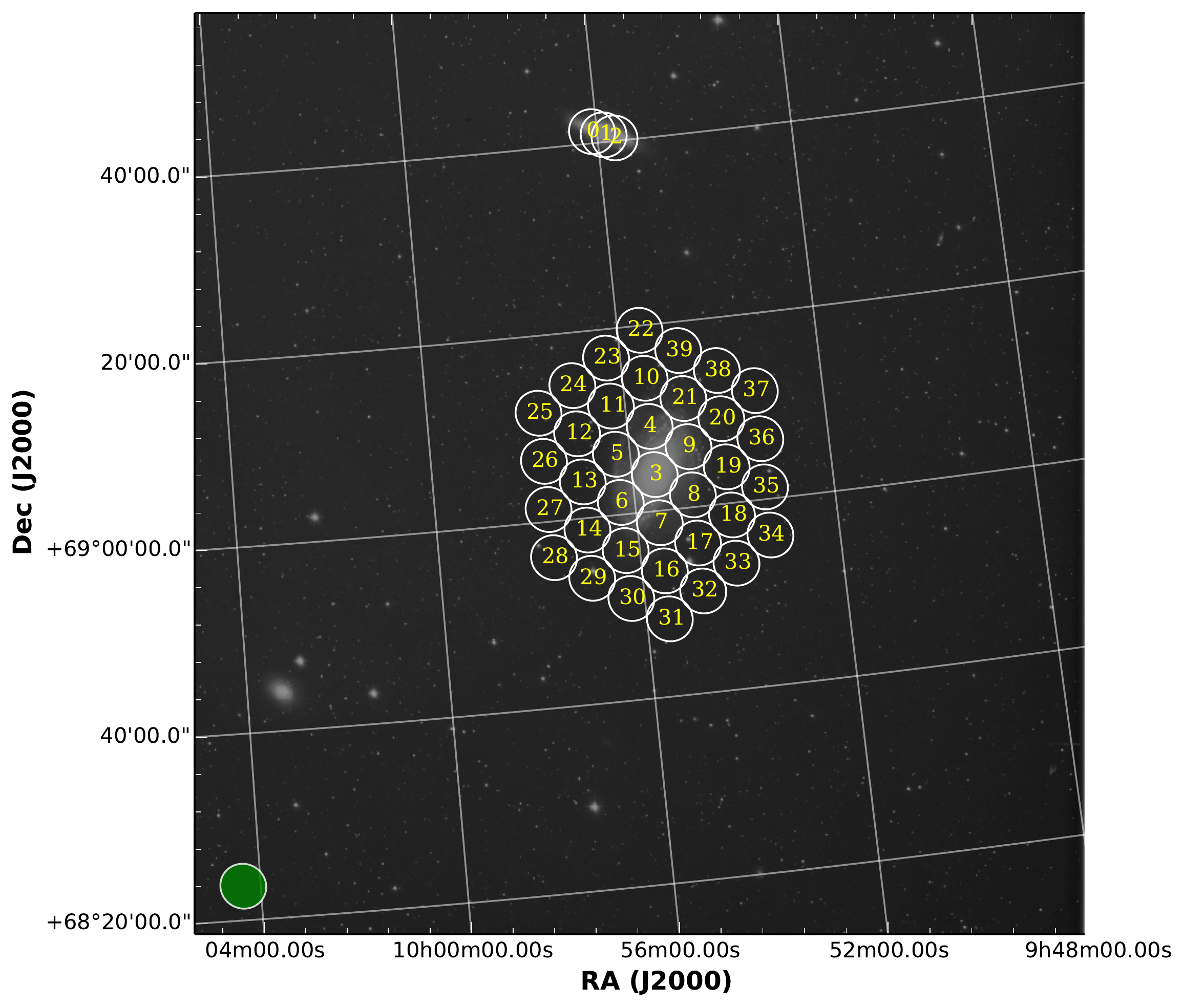}\label{m81m82}
}
\captionsetup{justification=centering}
\caption{Our targeted nearby galaxies:~\subref{m33} M33 -- beam 70 is missing due to one of the LOFAR processing-cluster nodes failing;~\subref{m81m82} M81/M82. The beam shape is shown in the bottom left corner of each panel.}
\label{Beams}
\end{figure}

We use the LOFAR HBA coherent core, coherently combining the central \bffinal{20 and 23} LOFAR core high-band antenna (HBA) stations \bffinal{for first and second observations, respectively}. These operate on the same central distributed clock and are thus the largest possible portion of the telescope that can be coherently combined in real time~\citep{Stappers-2011}. Figure~\ref{Beams} shows the tiling that was used, consisting of multiple rings of TABs, with radius successively increasing by 0.075 degrees. We formed five such rings for M33 during the first observation and three rings for M81 during the second observation. This resulted in the usage of 90 beams for M33 galaxy and 37 beams for M81 galaxy. Galaxy M82 was covered using three beams. Each observation was four hours long. LOFAR survey pointings as well as time integration and number of beams are demonstrated in Table~\ref{Obs}. 

Both observations were made with 68.5\,\mbox{MHz} of bandwidth around 146\,MHz. That is the maximum bandwidth available when covering the entire galaxy with tied-array beams, thus achieving highest possible sensitivity. Further observational characteristics are listed in Table~\ref{Setup}. The recorded 32-bit data were reduced to 8 bit and cleaned from RFI with the standard pulsar pipeline~\citep{Alexov-2010}.
 
\begin{table}[t!]
\centering
\caption{LOFAR Observational setup.}
\scalebox{0.7}{
\begin{tabular}{l r}
\hline\hline
\bf2ref{Parameter} & \bf2ref{Value} \\
\hline
Observational dates: & March 10 and May 9, 2015 \\
Telescope: & LOFAR \\
Receiver: & HBA \\
Backend: & COBALT \\
Number of tied-array beams \\
$-$ First observation: & 90 \\
$-$ Second observation: & 40 \\
Polarisations/beam: & 2 \\
Central frequency: & 146\,MHz \\
Frequency bandwidth: & 68.5\,MHz \\
Frequency channels: & 11232 \\
Sample time: & 1310\,$\mu$s \\
Integration time: & 14400\,s \\
\hline
\end{tabular}
}
\label{Setup}
\end{table}

\section{Data analysis}\label{sec3}

Through the LOFAR long term archive~\citep[LTA,][]{Renting-2011}, data were transferred to the Dutch national supercomputer Cartesius\footnote{\url{https://userinfo.surfsara.nl/systems/cartesius}}. There, we performed periodicity and single-pulse searches using~\texttt{PRESTO}~\citep{Ransom-2001}\bfref{, over the course of about 350,000 core-hours of Cartesius compute time.}

\subsection{Periodicity search}

Our search for periodic pulses was carried out independently and in parallel for each LOFAR beam. \bfref{All three galaxies have high inclinations: M33~\citep[55$^{\circ}$,][]{Hodge-2011} and M81~\citep[32$^{\circ}$,][]{Immler-2001}, M82~\citep[77$^{\circ}$,][]{Mayya-2005}, which was one of the selection and ranking criteria~\citep{Leeuwen-2010}.  The DM contributions from the Milky Way in their direction are about 50.1, 40.9, and 41.3\,pc\,cm$^{-3}$, respectively. Data were dedispersed up to a high DM of 1000\,pc\,cm$^{-3}$ DM grid, to account for potential giant molecular clouds situated in between, to allow for the errors on the free-electron models, and to remain sensitive to any low-frequency FRBs.} We erased particular RFI-affected frequencies, identified both per beam (both narrow band and at zero DM), and for general LOFAR signals. Our aim in this search is to detect the brightest sources in the target galaxies; these are most likely normal or young pulsars, not recycled MSPs. We thus focused on the somewhat slower, non-recycled section of the search parameter space. To limit processing time we started the search with a sample time 2.6 ms. Each TAB required $\approx 4\times10^4$ of DM steps to optimally correct for the possible dispersion smearing. Higher DMs were combined with successive downsampling in time. We then Fourier transformed the dedispersed time series and generally applied a periodicity search without attempting any corrections for binary acceleration. Finally, we folded the data up to 200 best candidates per beam, for subsequent manual inspection.

We first tested our search pipeline on two known pulsars that are characterised in Table~\ref{test}. Both these pulsars were found successfully. However, despite many promising candidates with high DM (around 600\,pc\,cm$^{-3}$) and small spin periods (around 100 ms), the main search revealed no new, persuasive extragalactic pulsars.

\begin{table}
\centering
\caption{Test pulsars found with LOFAR as a part of testing the search pipeline. \bfref{Tabulated are: integration time, number of used subbands, pulse period, DM, pulse duty cycle at 50\% of peak, expected flux density at 150\,MHz from the ATNF catalogue, and measured peak S/N.}}
\scalebox{0.68}{
\begin{tabular}{c c c c c c c c}
\hline\hline
Pulsar & $t_\textrm{int}$ & subbands & $P$ & DM & $w_{50} / P$ & $S_\mathrm{150}$ & $S/N_\textrm{peak}$ \\
 & (s) & & (ms) & (pc\,cm$^{-3}$) & & (Jy) & \\
\hline
PSR J0332+5434 & 260 & 128 & 714.501 & 26.77 & 9.2$\times10^{-3}$  & 8.8 & 99.21 \\
PSR J0953+0755 & 600 & 288 & 253.063 & 2.969 & 3.7$\times10^{-2}$ & 2.3 & 63.92 \\
\hline
\end{tabular}
}
\label{test}
\end{table}

\subsection{Single-pulse search}

\begin{figure}[t!]
\centering\subfigure[pulse $\#$1]{
\includegraphics[width=0.48\linewidth]{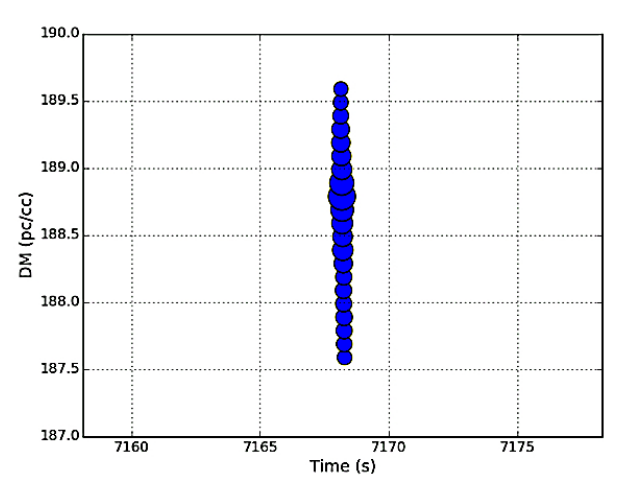}\label{m33_pulse1}
}\vfill
\centering\subfigure[pulse $\#$2]{
\includegraphics[width=0.48\linewidth]{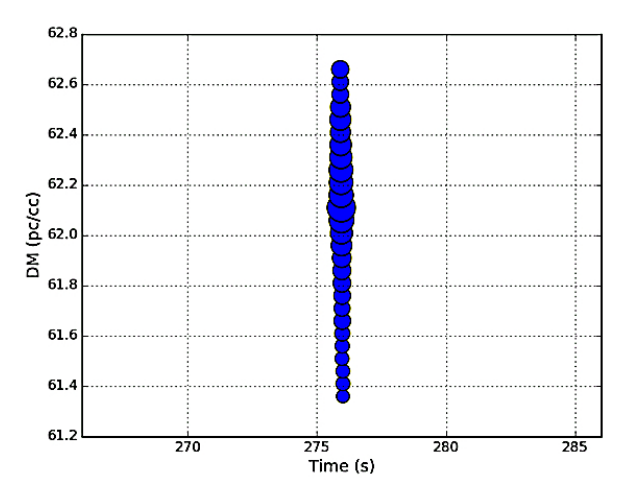}\label{m81m82_pulse1}
}\hfill
\centering\subfigure[pulse $\#$3]{
\includegraphics[width=0.48\linewidth]{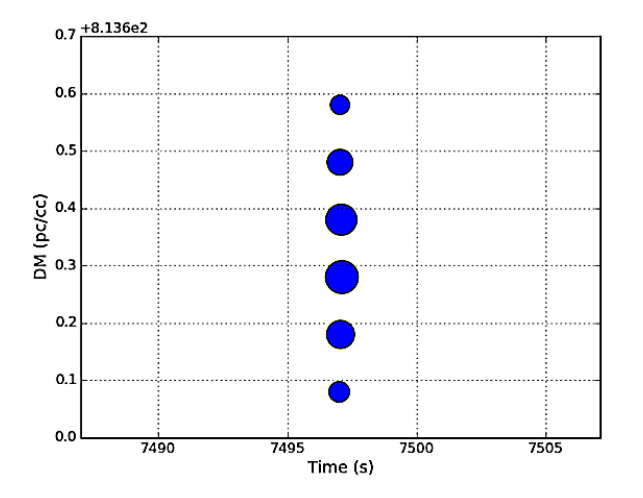}\label{m81m82_pulse2}
}\hfill
\captionsetup{justification=centering}
\caption{Typical DM versus time behaviour for aperiodic candidates -- S/N reduces either side of the peak due to the incorrect DM value:~\subref{m33_pulse1} for M33 galaxy;~\subref{m81m82_pulse1} and~\subref{m81m82_pulse2} for M81 and M82 galaxies \bf2ref{(the lack of characteristic points at higher DMs is due to the downsampling)}.}
\label{Pulses}
\end{figure}

\begin{figure}[t!]
\centering\subfigure[DM = 27.07\,pc\,cm$^{-3}$]{
\includegraphics[width=0.75\linewidth]{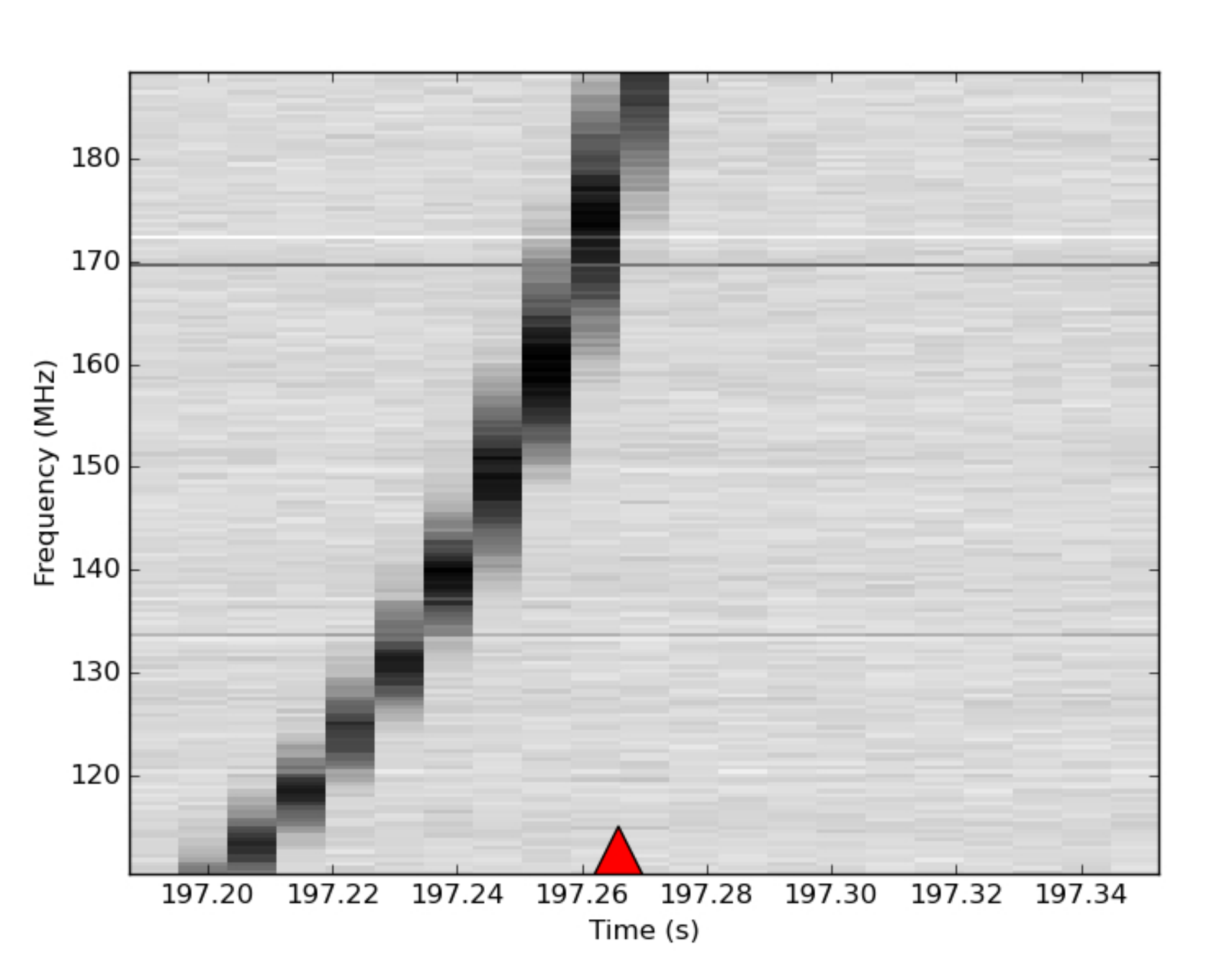}\label{DM27.07}
}\vfill
\centering\subfigure[DM = 26.77\,pc\,cm$^{-3}$]{
\includegraphics[width=0.75\linewidth]{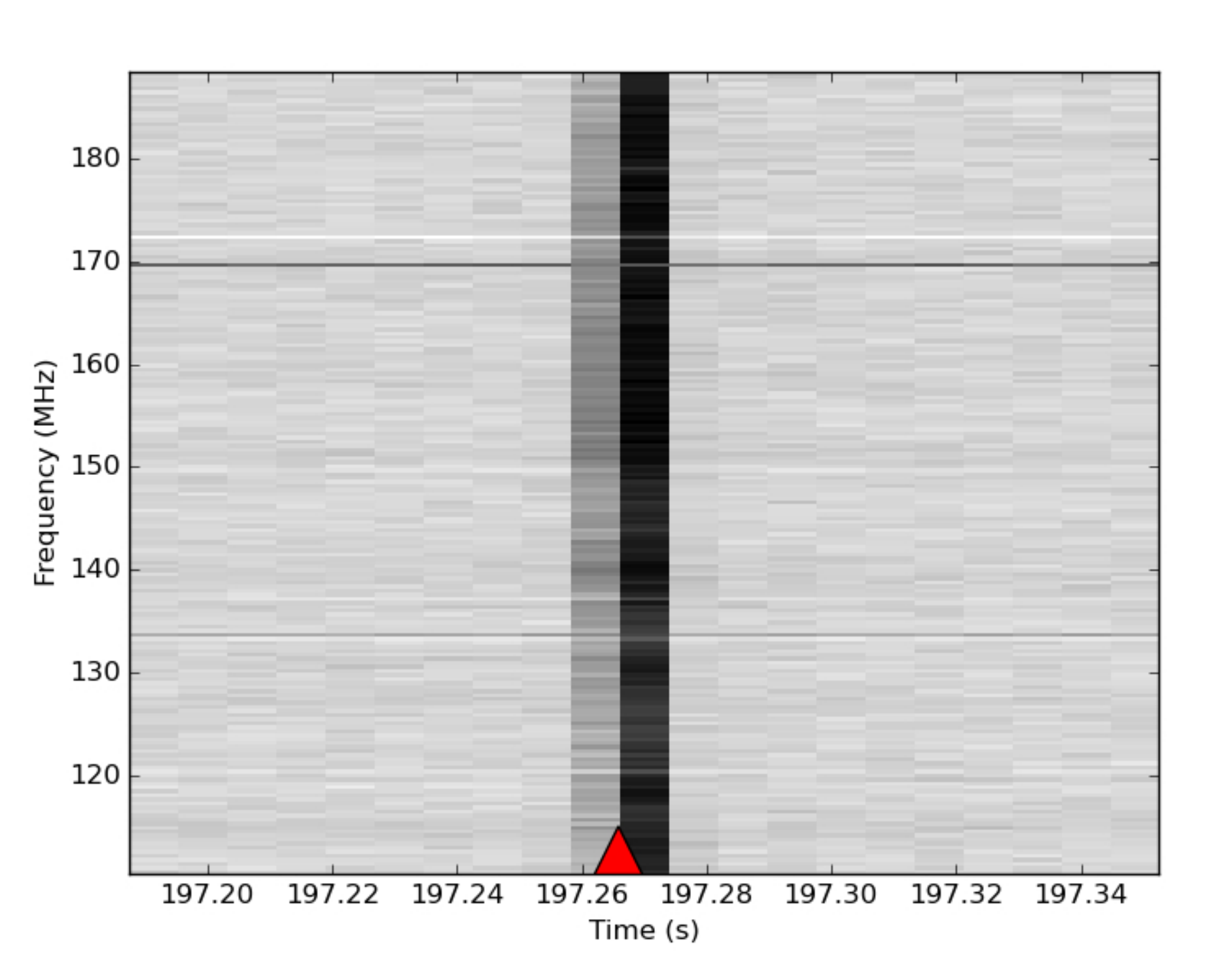}\label{DM26.77}
}\vfill
\captionsetup{justification=centering}
\caption{Example of the `waterfall' frequency versus time representation for the test pulsar PSR J0332+5434:~\subref{DM27.07} with an incorrect DM;~\subref{DM26.77} with a correct DM. \bf2ref{The red triangle denotes the expected position of the dispersion-corrected pulse signal on time series.}}
\label{Waterfall}
\end{figure}

To find sporadic sources, we started from dedispersed time series that we obtained earlier during the periodicity search step. We then searched for non-periodic pulses via boxcar function matched-filtering~\citep{Ransom-2001} and grouped all found pulses from all LOFAR TABs into a single database to simplify their subsequent characterisation~\citep[Michilli et al., 2016, in prep., see also ][for analogous approach]{Deneva-2016}. Both test pulsars were also successfully found with a single-pulse search routine and identified on the dedispersed raw data. Next, we ranked all the pulses from the database by the signal-to-noise ratio ($S/N_\mathrm{min}\geq$ 8), their extragalactic origin (DM $\geq$ DM$_z$, where DM$_z$ = 25 pc cm$^{-3}$ is the azimuthal Galactic DM contribution derived from the ne2001 model\footnote{\url{http://www.nrl.navy.mil/rsd/RORF/ne2001/}}~\citep{Cordes-2002} scaled to LOFAR frequencies) as well as their duration ($W\leq50$ ms).  Similar criteria were used by~\citet{Petroff-2015} in their search for extragalactic bursts. We looked for both repeated pulses at similar DM, as well as for completely single bursts. The main criteria for the pulse being genuine was a smooth decrease of S/N on both sides of the DM scale, see Fig.~\ref{Pulses}. The most promising candidates, both single and repeated, were then inspected directly in frequency versus time `waterfall' plots made from the raw data, see Fig.~\ref{Waterfall}. Most candidates turned out to be narrow-band RFI that remained after TAB RFI masking. In the end, no reliable astrophysical single pulses were found.

\section{Search results and implications}\label{sec4}

The search did not result in any new pulsars or transients from M33, M81 or M82. \bfref{However, not all pulsars are beamed our way. The beaming fraction~\citep{Smith-1969} relates to the chance of receiving radio emission from all the pulsars of a particular class. For young pulsars that might be born in other galaxies with typical periods of 100\,ms, only about 50\% are beamed toward Earth~\citep{Lorimer-2008}. We provide sensitivity estimates only for those extragalactic pulsars that can be potentially detected.} Below we therefore derive, for each, the limit on its pseudo-luminosity.

The periodic-search minimum detectable flux for a periodicity search~\citep[\bffinal{ps,}][]{Bhattacharya-1998}
\begin{equation}
S_{\mbox{\footnotesize{min}, ps}} = \beta\frac{T_\mathrm{sys}}{G\sqrt{n_\mathrm{p}\,\Delta\nu\,t_\mathrm{int}}} \times S/N_\mathrm{min} \times \sqrt{\frac{W}{P-W}}.
\end{equation}
Here $\beta$ is a digitisation factor (normally $\beta\gtrsim1$), $T_{sys}$ is the system temperature ($\mbox{K}$), $G$ is the telescope gain ($\mbox{K/Jy}$), $\Delta\nu$ is the bandwidth ($\mbox{Hz}$), and $t_\mathrm{int}$ is the observation dwell time ($\mbox{s}$). $S/N_\mathrm{min}$ corresponds to a pulse signal-to-noise threshold \bffinal{($\bmfinal{S/N_\mathrm{min}=10}$ in our case)}, and the final term quantifies how the sensitivity increases when the pulse flux is concentrated in a short pulse width $W$ every period $P$. \bffinal{The system temperature  and the LOFAR Core gain were estimated using \bffinal{the} Hamaker-Carozzi beam model with 50\% systematic uncertainties~\citep{Kondratiev-2015}: $\bmfinal{T_\mathrm{sys} = (7.4\pm1.1)\times10^2}$\,K, $\bmfinal{G = A_\mathrm{eff} / (2 \cdot k) \simeq4\pm2}$\,K/Jy. Here $\bmfinal{A_\mathrm{eff} = a_\mathrm{eff}\times N_\textrm{stations}^{\gamma} \times (1 - f_\textrm{bad})}$ is a total effective area, $\bmfinal{a_\mathrm{eff}}$ is a 48-tile HBA station effective area, $N_\textrm{stations}$ is the number of HBA stations (20 for M33 and 23 for M81/M82), $\bmfinal{\gamma=0.85}$ is the coherence factor, $\bmfinal{f_\textrm{bad}\simeq5\%}$ is the fraction of broken tiles, and $\bmfinal{k}$ is Boltzmann constant.} We obtained a noise rms flux density of \bffinal{$\bmfinal{0.15\pm0.08}$\,mJy for M33 and $\bmfinal{0.13\pm0.07}$\,mJy for M81/M82}. We next determine the expected observed pulse widths $W$ as a function of \bffinal{pulse} period $P$ and dispersion measure DM. We derive the intrinsic width $w_\mathrm{av}$ by applying an ATNF-averaged pulse duty cycle $\bmfinal{\langle w_\mathrm{50}/P \rangle \simeq 0.05}$ to the period $P$. We next include instrumental broadening effects: the DM stepsize smearing $w_\mathrm{dm}$, sub-band stepsize smearing $w_\mathrm{sub}$, and intra-channel smearing $w_\mathrm{chan}$\bfref{, and LOFAR sampling time constrain $w_\mathrm{samp}$}. Scattering may also severely increase the observed pulse width, and thus hinder detections at low frequencies. According to the ne2001 model, the maximum Galactic scatter broadening limits $w_\mathrm{scatter}$ in the direction to M33 and M81/M82 galaxies are 1.3\,ms and 0.9\,ms, respectively. We assume \bfref{two scattering screens (one in our Galaxy, and another in the host galaxy) -- by assuming similar structure of both, we thus double the resulting scatter broadening times.} Furthermore, each LOFAR sensitivity curve depends on the DM searched -- we show curves for DM = 0, 10, 100, and 1000\,pc\,cm$^{-3}$.

\begin{figure*}[t!]
\subfigure[Our Galaxy]{
\includegraphics[width=0.5\linewidth]{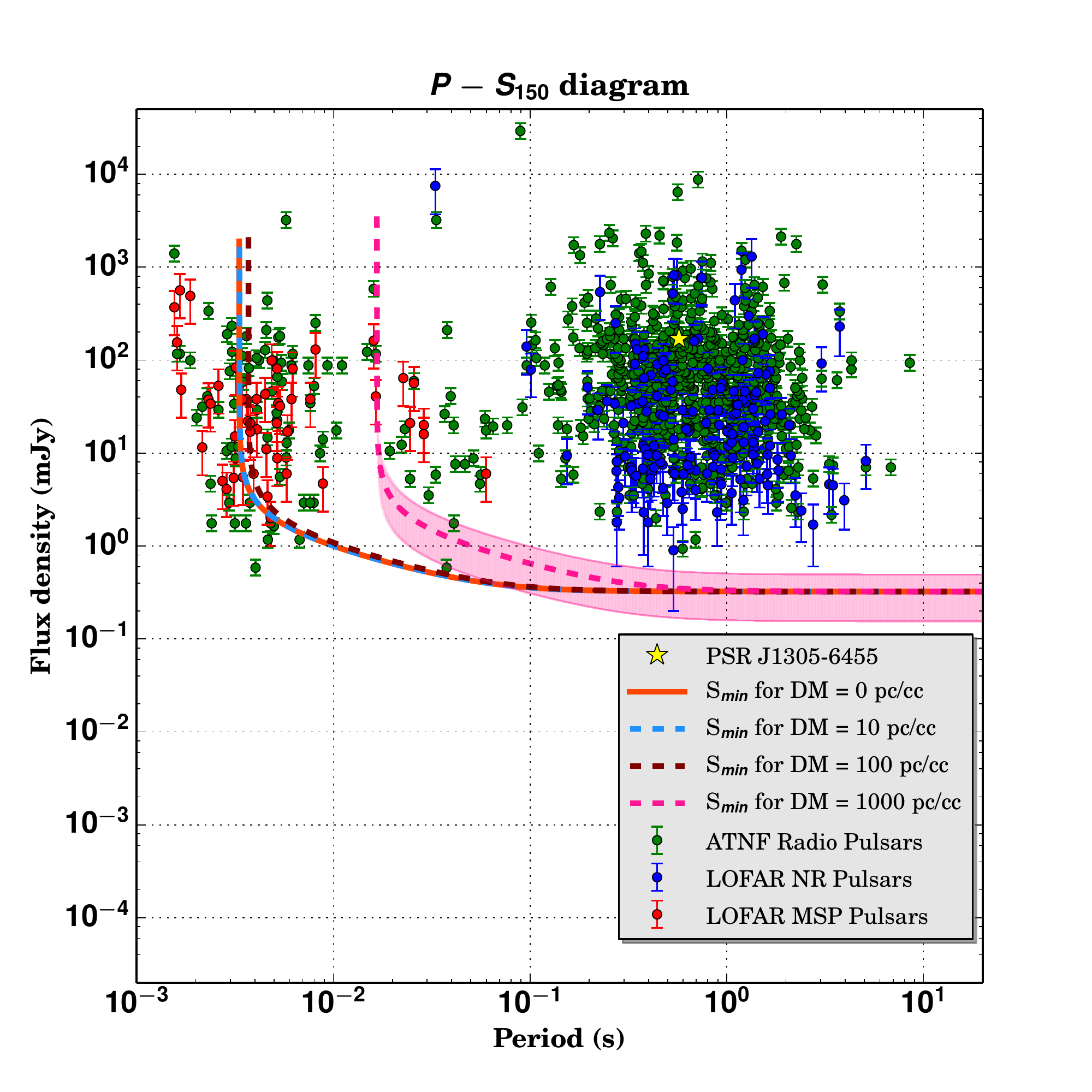}\label{galaxy}
}
\subfigure[Targeted galaxies]{
\includegraphics[width=0.5\linewidth]{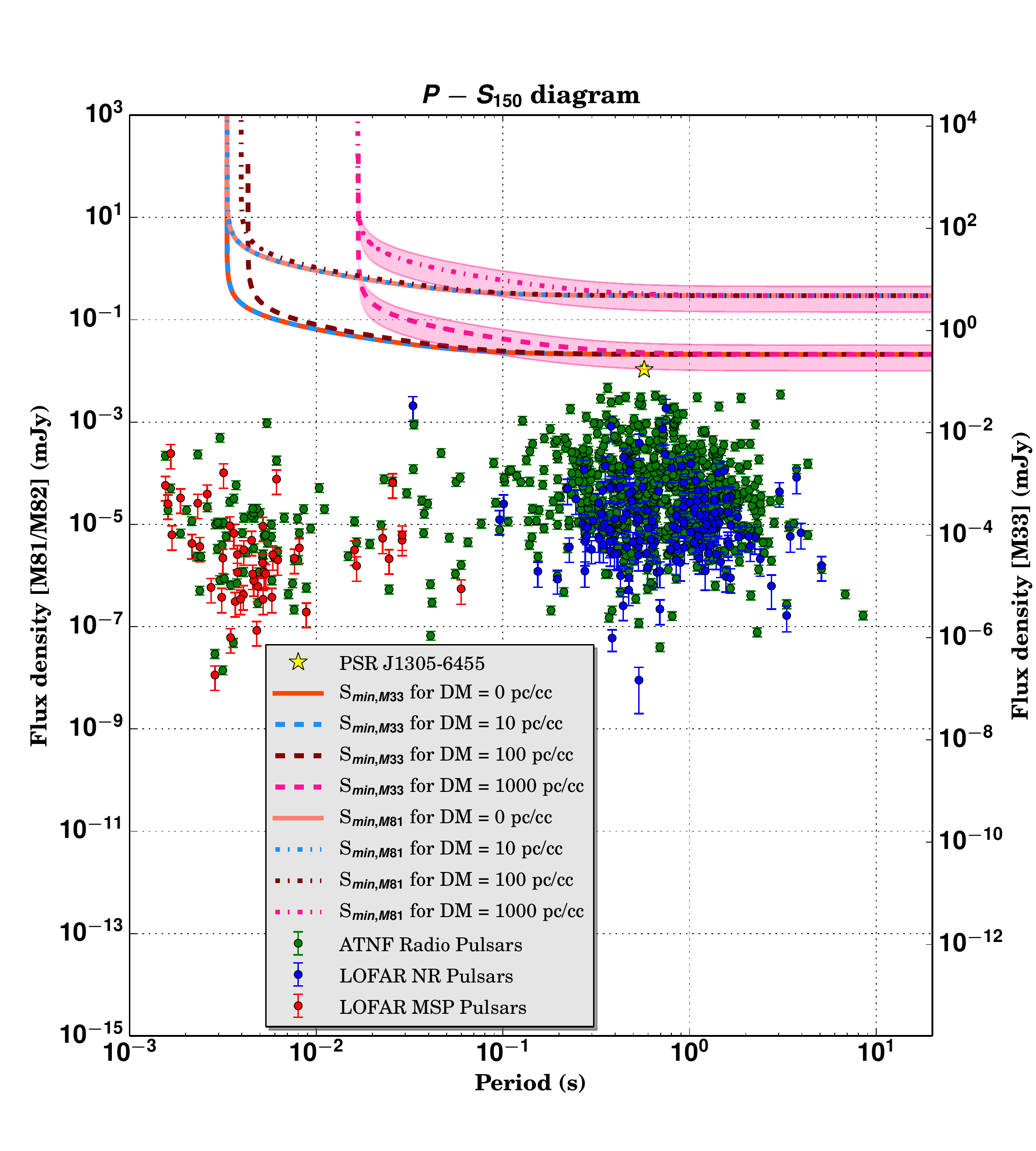}\label{ngalaxy}
}
\caption{\bffinal{Diagrams of flux density at 150\,MHz $S_{150}$ versus period ${P}$. Included are all the ATNF catalogue pulsars that have reported periods and flux densities. The LOFAR sensitivity curves for DM = 0, 10, 100, 1000\,pc\,cm$^{-3}$ are drawn, with the error contour for the most realistic DM for M33/M81/M82, at 1000\,pc\,cm$^{-3}$. The most-luminous Galactic pulsar, PSR J1305-6455, is denoted with a star. Panel \subref{galaxy} shows that our survey was sensitive enough to detect almost the entire Galactic pulsar population, had they been in the field of view. In \subref{ngalaxy} we illustrate our sensitivity to these same pulsars at the actual distance of M82/M81 (left y-axis) and M33 (right y-axis).}}
\label{gal_flux}
\end{figure*}

For a given DM, the broadening width is thus only dependent on the pulse period:
\begin{equation*}
W(P) = \sqrt{w_\mathrm{av}^2(P) + w_\mathrm{dm}^2 + w_\mathrm{sub}^2 + w_\mathrm{chan}^2 + w_\mathrm{samp}^2 + \left(2\,w_\mathrm{scatter}\right)^2}
\end{equation*}
which allows for the derivation of the sensitivity
\begin{equation}
S_{\mbox{\footnotesize{min}, ps}} \approx \bmfinal{0.2\pm0.1} \times 10 \times \sqrt{\frac{W(P)}{P-W(P)}}\,\mbox{mJy}.
\end{equation}
In addition to the periodicity sensitivity derived above, the minimum detectable flux for a single pulse search~\citep[\bffinal{sps,}][]{Cordes-2002}
\begin{equation}
\begin{split}
S_{\mbox{\footnotesize{min}, sps}} = \frac{T_\mathrm{sys}}{G\sqrt{n_\mathrm{p}\,\Delta\nu\,W}} \times S/N_\mathrm{min} = \\ = \frac{T_\mathrm{sys}}{G\sqrt{n_\mathrm{p}\,\Delta\nu\,t_\mathrm{int}}} \times S/N_\mathrm{min} \times \sqrt{\frac{t_\mathrm{int}}{W}}.
\end{split}
\end{equation}
Given the noise rms flux density limit and an upper limit on single pulses width ($0.05\,\mbox{s}$), we estimate the sensitivity
\begin{equation}
S_{\mbox{\footnotesize{min}, sps}} \approx \bmfinal{0.2\pm0.1} \times 8 \times 0.5\cdot10^3\,\mbox{mJy} \approx \bmfinal{0.8\pm0.4}\,\mbox{Jy}.
\end{equation}
To compare our limits to the known Galactic population, we can estimate how \bf2ref{bright} that population would appear to LOFAR, if placed in the targeted galaxies. For this purpose we first extracted from the ATNF pulsar archive all known pulsars flux densities at 400 MHz, which we scaled down to the LOFAR 150 MHz central frequency using a spectral index $\alpha = -1.8\pm0.2$~\citep{Maron-2000}. For reference we also added the millisecond~\citep{Kondratiev-2015} and non-recycled~\citep{Bilous-2015} pulsars that were previously detected with LOFAR, to our sensitivity plots. We see that our LOFAR survey observation is sensitive enough to catch most actual Galactic pulses, \bffinal{and only} fails to find \bffinal{the fastest} millisecond pulsars at high DMs, that is where \bfref{the average pulse width is larger than the initial sampling time ($\bmfinal{w_{50} \geq t_\mathrm{samp}}$) or} the broadening width is larger than the pulse period itself ($W \geq P$, Fig.~\ref{galaxy}). We note that the fact that many of these were detected by~\citet{Kondratiev-2015} is explained by their usage of for example, coherent dedispersion and folding on known ephemerides, much improving sensitivity over our blind-search filterbanked pipeline.

When referring to nearby galaxies, scaling with distance squared quickly diminishes the pulsar flux density. To estimate extragalactic pulsars fluxes, we have placed all ATNF catalogue pulsars
in the roughly equidistant pair M81/M82 and in M33 (Fig.~\ref{ngalaxy}). We compare our sensitivity curves with the M81/M82/M33 analogue of the very bright, distant ($d \simeq 30\,\mbox{kpc}$) Galactic pulsar J1305-6455 ($P \simeq 0.57\,\mbox{s}, S_\textrm{150, Galaxy} \simeq 0.17\,\mbox{Jy}$): we find $S_\textrm{150, M33} \simeq 0.17\,\mbox{mJy}$, $S_\textrm{150, M81/M82} \simeq 0.01\,\mbox{mJy}$. \bffinal{Our LOFAR sensitivity limit for this pulsar at} DM=1000\,pc\,cm$^{-3}$ is about \bffinal{0.53\,mJy} \bffinal{(Fig.~\ref{galaxy})}. We see that for M81/M82 the difference between the LOFAR sensitivity and the flux density of the brightest Galactic analogue pulsar is a little over \bffinal{an order of magnitude}. \bffinal{For M33, LOFAR is within a factor of a few of being able to detect the brightest Galactic analogue. With only slightly better sensitivity pulsar populations similar to that of the Galaxy could be resolved in M33.}

\section{Conclusions}\label{sec5}

We have conducted a deep LOFAR search for radio pulsars and  other time-domain transients in nearby galaxies M33, M81, and M82 with the highest currently-possible sensitivity at low frequencies. Using 130 LOFAR beams in total, we have searched up to DMs of 1000\,pc\,cm$^{-3}$ starting with 2.6 ms sampling time, four hours integration time. The detection of the two known test pulsars validated our search pipeline. We did, however, not detect any convincing new sources. We therefore established upper limits to the tip of the luminosity distributions on our target galaxies. \bffinal{Compared to the Milky Way population, we conclude there are no extragalactic pulsars brighter by only a factor of a few in M33, and by an order of magnitude in M81 or M82, shining our way.}

\begin{acknowledgements}
KM would like to thank E. Orru for scheduling the LOFAR observations, D. Michilli for providing single-pulse analysis tools, \bffinal{and F. Crawford for a valuable check of our calculations}. \bfref{We also thank the referee for useful comments that clarified the paper.}
LOFAR, the low frequency array designed and constructed by ASTRON, has facilities in several countries, that are owned by various parties (each with their own funding sources), and that are collectively operated by the International LOFAR Telescope (ILT) foundation under a joint scientific policy.
The research leading to these results has received funding from the European Research Council  under the European Union's Seventh Framework Programme (FP/2007-2013) / ERC Grant Agreement n. 617199, and from the Netherlands Research School for Astronomy (NOVA4-ARTS).
This work was carried out on the Dutch national e-infrastructure with the support of SURF Cooperative. Computing time was provided by NWO Physical Sciences.
\end{acknowledgements}


\bibliography{Ref}{}

\begin{thebibliography}{}
\providecommand\natexlab[1]{#1}
\providecommand\JournalTitle[1]{#1}

\bibitem[{{Alexov} {et~al.}(2010){Alexov}, {Hessels}, {Mol}, {Stappers}, \&
  {van Leeuwen}}]{Alexov-2010}
{Alexov}, A., {Hessels}, J., {Mol}, J.~D., {Stappers}, B., \& {van Leeuwen}, J.
  2010, in Astronomical Society of the Pacific Conference Series, Vol. 434,
  Astronomical Data Analysis Software and Systems XIX, ed. Y.~{Mizumoto}, K.-I.
  {Morita}, \& M.~{Ohishi}, 193

\bibitem[{{Bhat} {et~al.}(2011){Bhat}, {Cordes}, {Cox}, {Deneva}, {Hankins},
  {Lazio}, \& {McLaughlin}}]{Bhat-2011}
{Bhat}, N.~D.~R., {Cordes}, J.~M., {Cox}, P.~J., {et~al.} 2011,
  \href{http://dx.doi.org/10.1088/0004-637X/732/1/14}{\JournalTitle{\apj}, 732,
  14}

\bibitem[{{Bhattacharya}(1998)}]{Bhattacharya-1998}
{Bhattacharya}, D. 1998, in NATO Advanced Science Institutes (ASI) Series C,
  Vol. 515, NATO Advanced Science Institutes (ASI) Series C, ed. R.~{Buccheri},
  J.~{van Paradijs}, \& A.~{Alpar}, 103

\bibitem[{{Bilous} {et~al.}(2015){Bilous}, {Kondratiev}, {Kramer}, {Keane},
  {Hessels}, {Stappers}, {Malofeev}, {Sobey}, {Breton}, {Cooper}, {Falcke},
  {Karastergiou}, {Michilli}, {Os{\l}owski}, {Sanidas}, {ter Veen}, {van
  Leeuwen}, {Verbiest}, {Weltevrede}, {Zarka}, {Grie{\ss}meier}, {Serylak},
  {Bell}, {Broderick}, {Eisl{\"o}ffel}, {Markoff}, \&
  {Rowlinson}}]{Bilous-2015}
{Bilous}, A., {Kondratiev}, V., {Kramer}, M., {et~al.} 2015,
  \JournalTitle{ArXiv e-prints},
  \href{http://arxiv.org/abs/1511.01767}{{\sffamily arXiv:1511.01767
  [astro-ph.SR]}}

\bibitem[{{Camilo} \& {Rasio}(2005)}]{Camilo-2005}
{Camilo}, F., \& {Rasio}, F.~A. 2005, in Astronomical Society of the Pacific
  Conference Series, Vol. 328, Binary Radio Pulsars, ed. F.~A. {Rasio} \& I.~H.
  {Stairs}, 147

\bibitem[{{Coenen} {et~al.}(2014){Coenen}, {van Leeuwen}, {Hessels},
  {Stappers}, {Kondratiev}, {Alexov}, {Breton}, {Bilous}, {Cooper}, {Falcke},
  {Fallows}, {Gajjar}, {Grie{\ss}meier}, {Hassall}, {Karastergiou}, {Keane},
  {Kramer}, {Kuniyoshi}, {Noutsos}, {Os{\l}owski}, {Pilia}, {Serylak},
  {Schrijvers}, {Sobey}, {ter Veen}, {Verbiest}, {Weltevrede}, {Wijnholds},
  {Zagkouris}, {van Amesfoort}, {Anderson}, {Asgekar}, {Avruch}, {Bell},
  {Bentum}, {Bernardi}, {Best}, {Bonafede}, {Breitling}, {Broderick},
  {Br{\"u}ggen}, {Butcher}, {Ciardi}, {Corstanje}, {Deller}, {Duscha},
  {Eisl{\"o}ffel}, {Fender}, {Ferrari}, {Frieswijk}, {Garrett}, {de Gasperin},
  {de Geus}, {Gunst}, {Hamaker}, {Heald}, {Hoeft}, {van der Horst}, {Juette},
  {Kuper}, {Law}, {Mann}, {McFadden}, {McKay-Bukowski}, {McKean}, {Munk},
  {Orru}, {Paas}, {Pandey-Pommier}, {Polatidis}, {Reich}, {Renting},
  {R{\"o}ttgering}, {Rowlinson}, {Scaife}, {Schwarz}, {Sluman}, {Smirnov},
  {Swinbank}, {Tagger}, {Tang}, {Tasse}, {Thoudam}, {Toribio}, {Vermeulen},
  {Vocks}, {van Weeren}, {Wucknitz}, {Zarka}, \& {Zensus}}]{Coenen-2014}
{Coenen}, T., {van Leeuwen}, J., {Hessels}, J.~W.~T., {et~al.} 2014,
  \href{http://dx.doi.org/10.1051/0004-6361/201424495}{\JournalTitle{\aap},
  570, A60}

\bibitem[{{Cordes} \& {Lazio}(2002)}]{Cordes-2002}
{Cordes}, J.~M., \& {Lazio}, T.~J.~W. 2002, \JournalTitle{ArXiv Astrophysics
  e-prints}, \href{http://arxiv.org/abs/astro-ph/0207156}{{\sffamily
  astro-ph/0207156}}

\bibitem[{{Cordes} \& {McLaughlin}(2003)}]{Cordes-2003}
{Cordes}, J.~M., \& {McLaughlin}, M.~A. 2003,
  \href{http://dx.doi.org/10.1086/378231}{\JournalTitle{\apj}, 596, 1142}

\bibitem[{{Crawford} {et~al.}(2001){Crawford}, {Kaspi}, {Manchester}, {Lyne},
  {Camilo}, \& {D'Amico}}]{Crawford-2001}
{Crawford}, F., {Kaspi}, V.~M., {Manchester}, R.~N., {et~al.} 2001,
  \href{http://dx.doi.org/10.1086/320635}{\JournalTitle{\apj}, 553, 367}

\bibitem[{{Deneva} {et~al.}(2016){Deneva}, {Stovall}, {McLaughlin}, {Bagchi},
  {Bates}, {Freire}, {Martinez}, {Jenet}, \& {Garver-Daniels}}]{Deneva-2016}
{Deneva}, J.~S., {Stovall}, K., {McLaughlin}, M.~A., {et~al.} 2016,
  \href{http://dx.doi.org/10.3847/0004-637X/821/1/10}{\JournalTitle{\apj}, 821,
  10}

\bibitem[{{Hewish} {et~al.}(1968){Hewish}, {Bell}, {Pilkington}, {Scott}, \&
  {Collins}}]{Hewish-1968}
{Hewish}, A., {Bell}, S.~J., {Pilkington}, J.~D.~H., {Scott}, P.~F., \&
  {Collins}, R.~A. 1968,
  \href{http://dx.doi.org/10.1038/217709a0}{\JournalTitle{\nat}, 217, 709}

\bibitem[{Hodge(2011)}]{Hodge-2011}
Hodge, P. 2011, The Spiral Galaxy M33, Astrophysics and Space Science Library
  (Springer Netherlands)

\bibitem[{{Immler} \& {Wang}(2001)}]{Immler-2001}
{Immler}, S., \& {Wang}, Q.~D. 2001,
  \href{http://dx.doi.org/10.1086/321335}{\JournalTitle{\apj}, 554, 202}

\bibitem[{{Keane} {et~al.}(2011){Keane}, {Kramer}, {Lyne}, {Stappers}, \&
  {McLaughlin}}]{Keane-2011}
{Keane}, E.~F., {Kramer}, M., {Lyne}, A.~G., {Stappers}, B.~W., \&
  {McLaughlin}, M.~A. 2011,
  \href{http://dx.doi.org/10.1111/j.1365-2966.2011.18917.x}{\JournalTitle{\mnras},
  415, 3065}

\bibitem[{{Kondratiev} {et~al.}(2013){Kondratiev}, {Lorimer}, {McLaughlin}, \&
  {Ransom}}]{Vlad-2013}
{Kondratiev}, V., {Lorimer}, D., {McLaughlin}, M., \& {Ransom}, S. 2013,
  \href{http://dx.doi.org/10.1017/S1743921312024398}{in IAU Symposium, Vol.
  291, Neutron Stars and Pulsars: Challenges and Opportunities after 80 years,
  ed. J.~{van Leeuwen}}, 431

\bibitem[{{Kondratiev} {et~al.}(2015){Kondratiev}, {Verbiest}, {Hessels},
  {Bilous}, {Stappers}, {Kramer}, {Keane}, {Noutsos}, {Os{\l}owski}, {Breton},
  {Hassall}, {Alexov}, {Cooper}, {Falcke}, {Grie{\ss}meier}, {Karastergiou},
  {Kuniyoshi}, {Pilia}, {Sobey}, {ter Veen}, {Weltevrede}, {Bell}, {Broderick},
  {Corbel}, {Eisl{\"o}ffel}, {Markoff}, {Rowlinson}, {Swinbank}, {Wijers},
  {Wijnands}, \& {Zarka}}]{Kondratiev-2015}
{Kondratiev}, V.~I., {Verbiest}, J.~P.~W., {Hessels}, J.~W.~T., {et~al.} 2015,
  \JournalTitle{ArXiv e-prints},
  \href{http://arxiv.org/abs/1508.02948}{{\sffamily arXiv:1508.02948
  [astro-ph.HE]}}

\bibitem[{{Lorimer}(2008)}]{Lorimer-2008}
{Lorimer}, D.~R. 2008,
  \href{http://dx.doi.org/10.12942/lrr-2008-8}{\JournalTitle{Living Reviews in
  Relativity}, 11, 8}

\bibitem[{{Lorimer} \& {Kramer}(2004)}]{LorimerKramer-2004}
{Lorimer}, D.~R., \& {Kramer}, M. 2004, {Handbook of Pulsar Astronomy}
  (Cambridge University Press)

\bibitem[{{Manchester} {et~al.}(2005){Manchester}, {Hobbs}, {Teoh}, \&
  {Hobbs}}]{Manchester-2005}
{Manchester}, R.~N., {Hobbs}, G.~B., {Teoh}, A., \& {Hobbs}, M. 2005,
  \href{http://dx.doi.org/10.1086/428488}{\JournalTitle{\aj}, 129, 1993}

\bibitem[{{Maron} {et~al.}(2000){Maron}, {Kijak}, {Kramer}, \&
  {Wielebinski}}]{Maron-2000}
{Maron}, O., {Kijak}, J., {Kramer}, M., \& {Wielebinski}, R. 2000,
  \href{http://dx.doi.org/10.1051/aas:2000298}{\JournalTitle{\aaps}, 147, 195}

\bibitem[{{Mayya} {et~al.}(2005){Mayya}, {Carrasco}, \& {Luna}}]{Mayya-2005}
{Mayya}, Y.~D., {Carrasco}, L., \& {Luna}, A. 2005,
  \href{http://dx.doi.org/10.1086/432644}{\JournalTitle{\apjl}, 628, L33}

\bibitem[{{McLaughlin} \& {Cordes}(2003)}]{McLaughlin-2003}
{McLaughlin}, M.~A., \& {Cordes}, J.~M. 2003,
  \href{http://dx.doi.org/10.1086/378232}{\JournalTitle{\apj}, 596, 982}

\bibitem[{{McLaughlin} {et~al.}(2006){McLaughlin}, {Lyne}, {Lorimer}, {Kramer},
  {Faulkner}, {Manchester}, {Cordes}, {Camilo}, {Possenti}, {Stairs}, {Hobbs},
  {D'Amico}, {Burgay}, \& {O'Brien}}]{McLaughlin-2006}
{McLaughlin}, M.~A., {Lyne}, A.~G., {Lorimer}, D.~R., {et~al.} 2006,
  \href{http://dx.doi.org/10.1038/nature04440}{\JournalTitle{\nat}, 439, 817}

\bibitem[{{Noori} {et~al.}(2014){Noori}, {Roberts}, {Champion}, {McLaughlin},
  {Ransom}, \& {Ray}}]{Noori-2014}
{Noori}, H.~A., {Roberts}, M., {Champion}, D., {et~al.} 2014, in American
  Astronomical Society Meeting Abstracts, Vol. 223, American Astronomical
  Society Meeting Abstracts \#223, 153.15

\bibitem[{{Petroff} {et~al.}(2015){Petroff}, {Johnston}, {Keane}, {van
  Straten}, {Bailes}, {Barr}, {Barsdell}, {Burke-Spolaor}, {Caleb}, {Champion},
  {Flynn}, {Jameson}, {Kramer}, {Ng}, {Possenti}, \& {Stappers}}]{Petroff-2015}
{Petroff}, E., {Johnston}, S., {Keane}, E.~F., {et~al.} 2015,
  \JournalTitle{ArXiv e-prints},
  \href{http://arxiv.org/abs/1508.04884}{{\sffamily arXiv:1508.04884
  [astro-ph.HE]}}

\bibitem[{{Ransom}(2001)}]{Ransom-2001}
{Ransom}, S.~M. 2001, PhD thesis, Harvard University

\bibitem[{{Renting} \& {Holties}(2011)}]{Renting-2011}
{Renting}, G.~A., \& {Holties}, H.~A. 2011, in Astronomical Society of the
  Pacific Conference Series, Vol. 442, Astronomical Data Analysis Software and
  Systems XX, ed. I.~N. {Evans}, A.~{Accomazzi}, D.~J. {Mink}, \& A.~H. {Rots},
  49

\bibitem[{{Ridley} {et~al.}(2013){Ridley}, {Crawford}, {Lorimer}, {Bailey},
  {Madden}, {Anella}, \& {Chennamangalam}}]{Ridley-2013}
{Ridley}, J.~P., {Crawford}, F., {Lorimer}, D.~R., {et~al.} 2013,
  \href{http://dx.doi.org/10.1093/mnras/stt709}{\JournalTitle{\mnras}, 433,
  138}

\bibitem[{{Rubio-Herrera} \& {Maccarone}(2013)}]{RubioHerrera-2013}
{Rubio-Herrera}, E., \& {Maccarone}, T. 2013,
  \href{http://dx.doi.org/10.1017/S1743921312023307}{in IAU Symposium, Vol.
  291, IAU Symposium, ed. J.~{van Leeuwen}}, 111

\bibitem[{{Rubio-Herrera} {et~al.}(2013){Rubio-Herrera}, {Stappers}, {Hessels},
  \& {Braun}}]{Rubio-Herrera-2013}
{Rubio-Herrera}, E., {Stappers}, B.~W., {Hessels}, J.~W.~T., \& {Braun}, R.
  2013, \href{http://dx.doi.org/10.1093/mnras/sts205}{\JournalTitle{\mnras},
  428, 2857}

\bibitem[{{Smith}(1969)}]{Smith-1969}
{Smith}, F.~G. 1969,
  \href{http://dx.doi.org/10.1038/223934a0}{\JournalTitle{\nat}, 223, 934}

\bibitem[{{Spitler} {et~al.}(2016){Spitler}, {Scholz}, {Hessels}, {Bogdanov},
  {Brazier}, {Camilo}, {Chatterjee}, {Cordes}, {Crawford}, {Deneva}, {Ferdman},
  {Freire}, {Kaspi}, {Lazarus}, {Lynch}, {Madsen}, {McLaughlin}, {Patel},
  {Ransom}, {Seymour}, {Stairs}, {Stappers}, {van Leeuwen}, \&
  {Zhu}}]{Spitler-2016}
{Spitler}, L.~G., {Scholz}, P., {Hessels}, J.~W.~T., {et~al.} 2016,
  \href{http://dx.doi.org/10.1038/nature17168}{\JournalTitle{\nat}, 531, 202}

\bibitem[{{Stappers} {et~al.}(2011){Stappers}, {Hessels}, {Alexov}, {Anderson},
  {Coenen}, {Hassall}, {Karastergiou}, {Kondratiev}, {Kramer}, {van Leeuwen},
  {Mol}, {Noutsos}, {Romein}, {Weltevrede}, {Fender}, {Wijers}, {B{\"a}hren},
  {Bell}, {Broderick}, {Daw}, {Dhillon}, {Eisl{\"o}ffel}, {Falcke},
  {Griessmeier}, {Law}, {Markoff}, {Miller-Jones}, {Scheers}, {Spreeuw},
  {Swinbank}, {Ter Veen}, {Wise}, {Wucknitz}, {Zarka}, {Anderson}, {Asgekar},
  {Avruch}, {Beck}, {Bennema}, {Bentum}, {Best}, {Bregman}, {Brentjens}, {van
  de Brink}, {Broekema}, {Brouw}, {Br{\"u}ggen}, {de Bruyn}, {Butcher},
  {Ciardi}, {Conway}, {Dettmar}, {van Duin}, {van Enst}, {Garrett}, {Gerbers},
  {Grit}, {Gunst}, {van Haarlem}, {Hamaker}, {Heald}, {Hoeft}, {Holties},
  {Horneffer}, {Koopmans}, {Kuper}, {Loose}, {Maat}, {McKay-Bukowski},
  {McKean}, {Miley}, {Morganti}, {Nijboer}, {Noordam}, {Norden}, {Olofsson},
  {Pandey-Pommier}, {Polatidis}, {Reich}, {R{\"o}ttgering}, {Schoenmakers},
  {Sluman}, {Smirnov}, {Steinmetz}, {Sterks}, {Tagger}, {Tang}, {Vermeulen},
  {Vermaas}, {Vogt}, {de Vos}, {Wijnholds}, {Yatawatta}, \&
  {Zensus}}]{Stappers-2011}
{Stappers}, B.~W., {Hessels}, J.~W.~T., {Alexov}, A., {et~al.} 2011,
  \href{http://dx.doi.org/10.1051/0004-6361/201116681}{\JournalTitle{\aap},
  530, A80}

\bibitem[{{van Haarlem} {et~al.}(2013){van Haarlem}, {Wise}, {Gunst}, {Heald},
  {McKean}, {Hessels}, {de Bruyn}, {Nijboer}, {Swinbank}, {Fallows},
  {Brentjens}, {Nelles}, {Beck}, {Falcke}, {Fender}, {H{\"o}randel},
  {Koopmans}, {Mann}, {Miley}, {R{\"o}ttgering}, {Stappers}, {Wijers},
  {Zaroubi}, {van den Akker}, {Alexov}, {Anderson}, {Anderson}, {van Ardenne},
  {Arts}, {Asgekar}, {Avruch}, {Batejat}, {B{\"a}hren}, {Bell}, {Bell}, {van
  Bemmel}, {Bennema}, {Bentum}, {Bernardi}, {Best}, {B{\^i}rzan}, {Bonafede},
  {Boonstra}, {Braun}, {Bregman}, {Breitling}, {van de Brink}, {Broderick},
  {Broekema}, {Brouw}, {Br{\"u}ggen}, {Butcher}, {van Cappellen}, {Ciardi},
  {Coenen}, {Conway}, {Coolen}, {Corstanje}, {Damstra}, {Davies}, {Deller},
  {Dettmar}, {van Diepen}, {Dijkstra}, {Donker}, {Doorduin}, {Dromer}, {Drost},
  {van Duin}, {Eisl{\"o}ffel}, {van Enst}, {Ferrari}, {Frieswijk}, {Gankema},
  {Garrett}, {de Gasperin}, {Gerbers}, {de Geus}, {Grie{\ss}meier}, {Grit},
  {Gruppen}, {Hamaker}, {Hassall}, {Hoeft}, {Holties}, {Horneffer}, {van der
  Horst}, {van Houwelingen}, {Huijgen}, {Iacobelli}, {Intema}, {Jackson},
  {Jelic}, {de Jong}, {Juette}, {Kant}, {Karastergiou}, {Koers}, {Kollen},
  {Kondratiev}, {Kooistra}, {Koopman}, {Koster}, {Kuniyoshi}, {Kramer},
  {Kuper}, {Lambropoulos}, {Law}, {van Leeuwen}, {Lemaitre}, {Loose}, {Maat},
  {Macario}, {Markoff}, {Masters}, {McFadden}, {McKay-Bukowski}, {Meijering},
  {Meulman}, {Mevius}, {Middelberg}, {Millenaar}, {Miller-Jones}, {Mohan},
  {Mol}, {Morawietz}, {Morganti}, {Mulcahy}, {Mulder}, {Munk}, {Nieuwenhuis},
  {van Nieuwpoort}, {Noordam}, {Norden}, {Noutsos}, {Offringa}, {Olofsson},
  {Omar}, {Orr{\'u}}, {Overeem}, {Paas}, {Pandey-Pommier}, {Pandey}, {Pizzo},
  {Polatidis}, {Rafferty}, {Rawlings}, {Reich}, {de Reijer}, {Reitsma},
  {Renting}, {Riemers}, {Rol}, {Romein}, {Roosjen}, {Ruiter}, {Scaife}, {van
  der Schaaf}, {Scheers}, {Schellart}, {Schoenmakers}, {Schoonderbeek},
  {Serylak}, {Shulevski}, {Sluman}, {Smirnov}, {Sobey}, {Spreeuw}, {Steinmetz},
  {Sterks}, {Stiepel}, {Stuurwold}, {Tagger}, {Tang}, {Tasse}, {Thomas},
  {Thoudam}, {Toribio}, {van der Tol}, {Usov}, {van Veelen}, {van der Veen},
  {ter Veen}, {Verbiest}, {Vermeulen}, {Vermaas}, {Vocks}, {Vogt}, {de Vos},
  {van der Wal}, {van Weeren}, {Weggemans}, {Weltevrede}, {White}, {Wijnholds},
  {Wilhelmsson}, {Wucknitz}, {Yatawatta}, {Zarka}, {Zensus}, \& {van
  Zwieten}}]{Haarlem-2013}
{van Haarlem}, M.~P., {Wise}, M.~W., {Gunst}, A.~W., {et~al.} 2013,
  \href{http://dx.doi.org/10.1051/0004-6361/201220873}{\JournalTitle{\aap},
  556, A2}

\bibitem[{{van Leeuwen} \& {Stappers}(2010)}]{Leeuwen-2010}
{van Leeuwen}, J., \& {Stappers}, B.~W. 2010,
  \href{http://dx.doi.org/10.1051/0004-6361/200913121}{\JournalTitle{\aap},
  509, A7}

\end{thebibliography}
\bibliographystyle{yahapj}

\end{document}